\documentclass[letterpaper, 10 pt, conference]{ieeeconf}  

\IEEEoverridecommandlockouts                              

\usepackage{amsmath,graphicx}
\usepackage{booktabs}
\usepackage{color, colortbl}
\usepackage{multirow}
\usepackage{hyperref}

\usepackage[backend=biber,
    sorting=none,
    citestyle=numeric,
    style=numeric,
    natbib=true]{biblatex}
\renewcommand{\baselinestretch}{0.9577}

\newcolumntype{L}{>{\centering\arraybackslash}m{3cm}}

\newcommand\app[1]{\textit{SmartKC}}

\title{\LARGE \bf
Keratoconus Classifier for Smartphone-based Corneal Topographer
}

\vspace{-7.0mm}

\author{Siddhartha Gairola$^{1}$, Pallavi Joshi$^{2}$, Anand Balasubramaniam$^{2}$, Kaushik Murali$^{2}$, Nipun Kwatra$^{1}$, Mohit Jain$^{1}$
\thanks{$^{1}$ Microsoft Research, India
        {\tt\small \{t-sigai, nkwatra, mohja\} @microsoft.com}}%
\thanks{$^{2}$ Sankara Eye Hospital, Bengaluru, India}
}

\begin{document}

\maketitle
\thispagestyle{empty}
\pagestyle{empty}

\begin{abstract}

Keratoconus is a severe eye disease that leads to deformation of the cornea. It impacts people aged 10-25 years and is the leading cause of blindness in that demography. Corneal topography is the gold standard for keratoconus diagnosis. It is a non-invasive process performed using expensive and bulky medical devices called corneal topographers. This makes it inaccessible to large populations, especially in the Global South. Low-cost smartphone-based corneal topographers, such as \app{}, have been proposed to make keratoconus diagnosis accessible. Similar to medical-grade topographers, \app{} outputs curvature heatmaps and quantitative metrics that need to be evaluated by doctors for keratoconus diagnosis. An automatic scheme for evaluation of these heatmaps and quantitative values can play a crucial role in screening keratoconus in areas where doctors are not available. In this work, we propose a dual-head convolutional neural network (CNN) for classifying keratoconus on the heatmaps generated by \app{}. Since \app{} is a new device and only had a small dataset (114 samples), we developed a 2-stage transfer learning strategy---using
historical data collected from a medical-grade topographer and a subset of \app{} data---to satisfactorily train our network. This, combined with our domain-specific data augmentations, achieved a sensitivity of 91.3\% and a specificity of 94.2\%.

\end{abstract}
\section{Introduction}

Keratoconus is an eye disease that causes the cornea to become thin, leading to a conical shape. It results in blurred vision, irregular astigmatism, and partial/complete blindness. Studies have shown that keratoconus is highly prevalent in the Global South, \textit{e.g.}, in central India, 1 in 44 individuals were found to suffer from keratoconus~\cite{kc_prevalence_3}. 
The disease generally affects people between the ages of 10 and 25, and may progress slowly for years. Early diagnosis is crucial to provide timely treatment and prevent worsening of disease. 
The clinical gold standard to diagnose keratoconus is using a technique called corneal topography. This is done by a sophisticated medical device called corneal topographer. It measures the shape of the corneal surface and outputs curvature heatmaps, helping the doctor to diagnose keratoconus. Although highly accurate, these devices are expensive, bulky (non-portable), and require a trained technician to operate. 
Such factors limit accessibility and also make frequent mass screening of keratoconus hard in remote areas.

\begin{figure}[t]
\begin{center}
    \centering
    \includegraphics[width=0.75\linewidth]{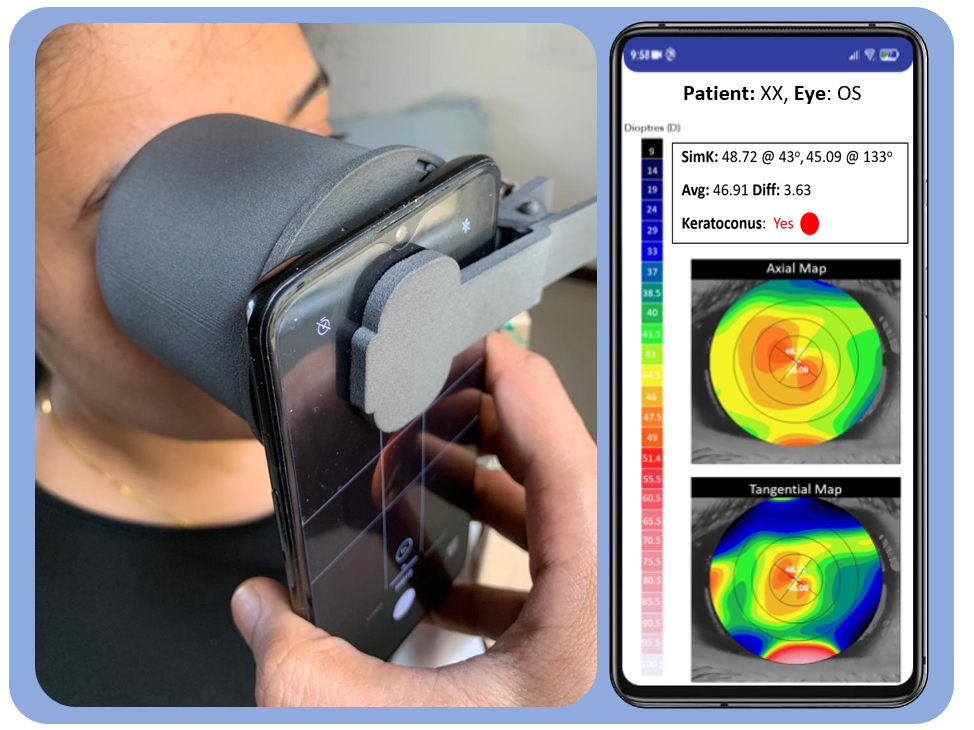}
\end{center}
\vspace{-5mm}
    \caption{Left: \app{}, a smartphone-based corneal topographer. Right: Outputs generated by \app{} (axial heatmap, tangential heatmap, sim-K values, and automatic keratoconus diagnosis by our proposed model).} 
    \label{fig:smartkc_system}
\vspace{-6mm}
\end{figure}

In our prior work, we proposed \app{}~\cite{smartkc} (Figure~\ref{fig:smartkc_system}), a low-cost smartphone-based corneal topographer that generates topography heatmaps similar to a medical-grade corneal topographer. Through a clinical study comprising of 101 eyes (57 participants) and evaluation by four ophthalmologists, we showed that \app{} achieved a sensitivity of 94.1\% and specificity of 100.0\%.
However, \app{}-generated heatmaps need to be evaluated by doctors for diagnosis, and with only 1 doctor for more than 1000 people in the Global South~\cite{who_doctors},
it is hard to obtain a doctor's evaluation for every keratoconus screening test.
Additionally, evaluations by doctors suffer from subjectivity. For instance, in \app{} evaluation, we found that for 30.7\% of eyes, at least one of the four doctors had a different diagnosis than the rest (this number was 42.6\% for Keratron device). 
Having an accurate automated method to detect keratoconus that works for low-cost topographers can help alleviate such issues and enable mass screening to identify keratoconus patients.

Recent works have demonstrated the efficacy of using deep neural networks for accurately detecting keratoconus 
using color-coded heatmaps obtained from clinical topographers, such as devices based on Optical Coherence Tomography~\cite{kamiya_bmj_2019}, Scheimpflug imaging~\cite{chen_bmj_2021, ajo_2020, jrs_2021}, and placido disc reflection~\cite{kuo_tvst_2020, iovs_2021}. However, such methods have been limited to medical-grade topographers. We extend this line of work to adapt it to heatmaps generated from the low-cost \app{} device.

In this work, we propose a deep CNN with two classification heads, that takes \app{}-generated heatmaps as input and provides keratoconus diagnosis as output. Since we had very few samples (114 eyes) from \app{}~\cite{smartkc}, to effectively train our network we propose a 2-stage transfer learning strategy:
(a) pre-training on ImageNet followed by fine-tuning on historical data (2110 samples) collected from Optikon Keratron, a medical-grade topographer, and (b)
further fine-tuning on 50\% of the \app{}-generated heatmaps.
This improved the performance significantly (sensitivity=82.1\%, specificity=85.6\%), compared to the standard fine-tuning on only \app{} samples (sensitivity=65.2\%, specificity=76.5\%).
On further analysis, we found that Keratron-generated heatmaps had fixed scaling and centering due to the stable head- and chest-rest setup, while \app{} heatmaps had variations owing its handheld nature.
Hence, we applied domain-specific augmentations to simulate similar behavior in the Keratron dataset, which helped us achieve a sensitivity of 91.3\% and specificity of 94.2\%.

The main contributions of our work are:
\begin{enumerate}
    \item a dual-head CNN-based keratoconus detection algorithm for \app{}, a smartphone-based topographer,
    \item a 2-stage transfer learning strategy and domain-specific augmentations for training with very few samples, and
    \item evaluation of our method on topography heatmaps from actual patients obtained by \app{} and also a medical-grade topographer (Keratron). 
\end{enumerate}

\section{Related Work}

Prior research has proposed automated methods using quantitative indices~\cite{machine_indices} (like Percentage Probability of Keratoconus (PPK), Cone Location and Magnitude Index~\cite{clmi}), statistical methods~\cite{auto_2000, ml_classifier_2016_1}, and traditional machine learning algorithms~\cite{iovs_2021, ml_classifiers_2016, nature_2020} (like logistic regression, K-nearest neighbour, clustering, decision trees, and random forests) to diagnose keratoconus from corneal topography data. However, with the advent of deep learning and its high performance on image classification tasks, deep learning based methods are being widely explored for keratoconus diagnosis and have been shown to be highly accurate~\cite{chen_bmj_2021, ajo_2020, kerato_detect_2019}. 


\citet{kerato_detect_2019} propose to use a CNN-based classifier for keratoconus detection, however they train and test their model only on synthetic eye data.
\citet{ajo_2020} investigate the performance of CNNs to classify corneal topography heatmaps into normal and keratoconus. They evaluate the model on 3000 samples from real patients obtained using Orbscan, a medical-grade topographer, and achieve classification accuracy of 99.3\%. Similarly, a ResNet152 based classification model has been proposed~\cite{kuo_tvst_2020}. 
They train and test their model on 354 samples, and achieve sensitivity and specificity of above 90\%.
\citet{chen_bmj_2021} use corneal tomography scans, with four type of heatmaps (axial, anterior elevation, posterior elevation, and pachymetry) per sample, and adopt a VGG16 model to learn a keratoconus detection classifier. They train their model on 1115 samples and test it on 279 samples, achieving a sensitivity of 98.5\% and specificity of 90.0\%.

These systems have achieved high performance for keratoconus detection, but have been limited to training and testing on heatmaps generated by medical-grade topographers. 
In spite of the progress made in developing low-cost portable topographers~\cite{smartkc, phone-kt-spie-dots, phone-kt-spie-sidepic, phone-kt-3D-print, phone-kt-ieee, lvpei-phone-kt, kt-masters-thesis}, there has been no work on adapting a model designed for data obtained from medical-grade topographers that is suitable for such low-cost alternatives. In our work, we aim to bridge this gap by proposing a CNN-based solution that accurately detects keratoconus from corneal topography heatmaps obtained using both, \app{} (a low-cost portable topographer) and Keratron (a medical-grade topographer).

\section{Method}

\smallskip\noindent\textbf{Dataset}: 
Our dataset is divided into two parts: 

\smallskip\noindent \textit{\app{}-data}: This an augmented version of the dataset used in~\cite{smartkc}, collected from 64 patients, with 114 eye samples obtained from both Keratron and \app{} device deployed at the Sankara Eye Hospital
in Bengaluru, India. It comprises of 68 non-keratoconus and 46 keratoconus eyes. The ground truth data for each eye was keratoconus classification, as diagnosed by a senior ophthalmologist at the hospital. This data was collected during Jun-Aug'21.

\smallskip\noindent \textit{Keratron-data}: Since 114 samples is small for training a deep learning model, we retrospectively downloaded anonymized data from the Keratron device database for all the patients who took the corneal topography examination at the hospital from April'08 to May'10. The dataset consists of 2110 samples (1637 non-keratoconus and 473 keratoconus). Each sample comprises of three images (axial heatmap, tangential heatmap, and placido ring), and three quantitative metrics (sim-K1, sim-K2, and PPK). The ground truth class (keratoconus vs non-keratoconus) was obtained based on the PPK-based classification\footnote{PPK$<20\%$ is non-keratoconus, $20\%\le$PPK$<45\%$ is suspect keratoconus, and $45\%\le$ PPK is keratoconus, as per Keratron user-manual.}. 
All the details of our study were approved by the hospital's Institutional Review Board.

\smallskip\noindent \textbf{Data Preprocessing}: All heatmaps are cropped 
and resized to a fixed resolution of $512\times512$. The heatmaps are then standardized by performing Z-normalization on each channel of the RGB image ($\Tilde{X}=(X-\mu)/\sigma$, where the mean $\mu$ and standard deviation $\sigma$ are computed for the entire dataset).

\begin{figure}[t]
\begin{center}
    \centering
    \includegraphics[width=1\linewidth]{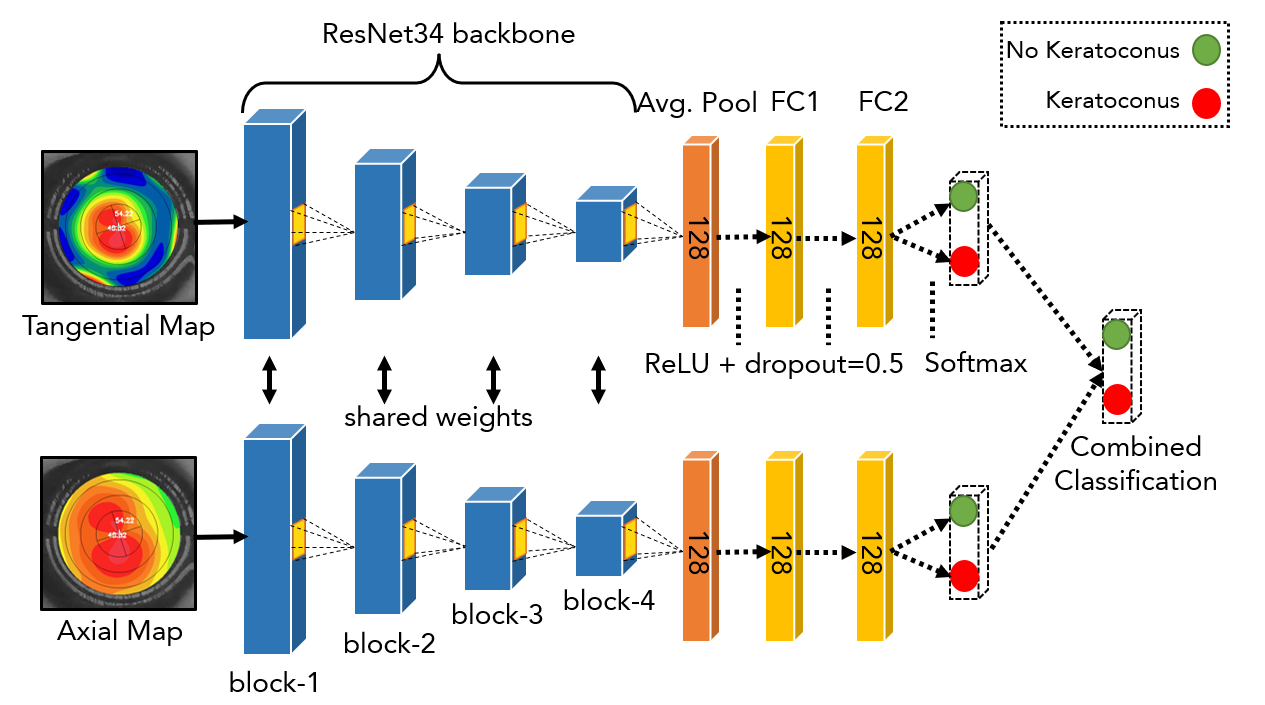}
\end{center}
\vspace{-5mm}
    \caption{\textit{Proposed network architecture.}} 
    \label{fig:overview}
\vspace{-4mm}
\end{figure}

\smallskip\noindent \textbf{Network Architecture}: Figure~\ref{fig:overview} illustrates the architecture of the proposed CNN. The network is organized into two branches---one each for axial and tangential heatmaps---with a shared convolutional backbone, followed by two distinct feed-forward classifiers, one for each branch. The shared backbone comprises of the convolutional layers from a ResNet34 model.
The classifiers consist of two \textit{128-d} fully connected (FC) layers with \textit{ReLU} activations. The last layer of each classifier applies \textit{softmax activation} to model class probabilities. \textit{Dropout} is added to each FC layer to prevent overfitting. The network is jointly trained via a \textit{weighted binary cross-entropy}\footnote{where the weights are inversely proportional to the number of samples in each class, to address class imbalance.} loss to minimize the loss for the two-class (keratoconus vs non-keratoconus) classification task.

\smallskip\noindent\textbf{Training Procedure}:
We followed a 2-stage transfer learning strategy to train our network. 
In stage-1, the network backbone is initialized with weights from a ResNet34 network pre-trained on ImageNet~\cite{ILSVRC15}, and then \textit{fine-tuned}
on the Keratron-data with a 90:10 train:validation split. This achieved good performance on the Keratron heatmaps, however failed on the \app{} heatmaps (see Tables~\ref{tab:result_comparison}, \ref{tab:ablations}).
On deeper analysis of the data, we noticed that as \app{} is a portable hand-held device, the heatmaps generated by it had variations in terms of zoom level, translation, and aspect ratio. In comparison, the heatmaps obtained from Keratron had minimal variations as they are captured using a stable head- and chin-rest setup. 
To handle this difference in image characteristics, we added data augmentations
(described below) and a stage-2 fine-tuning.
In stage-2, we fine-tuned the network on a subset (50\%, 57 samples) of the \app{}-data to ensure that the network learns these variations. This helped further improve the performance
(see Table~\ref{tab:ablations}).

\smallskip\noindent \textbf{Data Augmentation}: As the Keratron-data is relatively small (2110 samples) for training a deep neural network, we added \textit{domain-specific data augmentations} like horizontal-flip, rotation ($0^o-10^o$), scaling ($0.6x-1.4$x) and translation ($1-10\%$). Besides improving generalization of the network, this also helped in making the network robust to variations in the \app{} captured images.
Additionally, we also employ \textit{mixup data augmentations}~\cite{mixup}. 
To combine two randomly selected samples $x_i, x_j$ along with their labels $y_i, y_j$ in the training data, a new sample $\Tilde{x}$ and its label $\Tilde{y}$ is generated as follows using the mixup scheme: $\Tilde{x}=\lambda x_i + (1-\lambda)x_j$, $\Tilde{y}=\lambda y_i + (1-\lambda)y_j$, where $y_i, y_j$ are one-hot encoded class labels of the feature vectors $x_i, x_j$, and $\lambda \in [0,1]$ is a random number generated according to a Beta distribution~\cite{mixup}. Mixup also helps with addressing the data imbalance problem typical in medical datasets.


\section{Experiments and Results}

\smallskip\noindent\textbf{Implementation Details}: We train our network on a Tesla v100 GPU on a Microsoft Azure VM. We used the SGD optimizer with momentum of 0.9 and a batch size of 32, with a weighted random sampler to sample mini-batches uniformly from each class. For the stage-1 fine-tuning on Keratron-data, the network is trained for 200 epochs using a fixed learning rate of 1e-3 (took 73.5 minutes). For the stage-2 fine-tuning on \app{}-data, we start with a 1e-4 learning rate with linear decay for 100 epochs (took 3.5 minutes). The inference time for a single image was 0.3 seconds.
Note: During both stages of training, \textit{ResNet34 blocks 1-3} of the backbone were kept frozen.

\begin{table}[t]
\centering
 \resizebox{\linewidth}{!}{
\begin{tabular}{ |c | c | c c c|}
\toprule
 \textbf{Eval. Data} & \textbf{Model} & \textbf{Se} & \textbf{Sp} & \textbf{Acc}\\
\cmidrule(lr){1-1}
\cmidrule(lr){2-2}
\cmidrule(lr){3-5}

\multirow{3}{*}{\shortstack{SmartKC-data \\ (57 samples)}} & SVM & 80.4\% & 100.0\% & 92.1\% \\
&Dual-head CNN$^\dagger$ & 65.2\% & 76.5\% & 71.9\% \\
&Dual-head CNN* & 91.3\% & 94.2\% & 93.1\% \\
\hline
\multirow{3}{*}{\shortstack{Keratron-data \\ (114 samples)}} & PPK & 89.4\% & 94.2\% & 92.2\% \\
& SVM & 78.3\% & 95.5\% & 88.6\% \\
& Dual-head CNN$^\ddagger$ & 94.7\% & 93.4\% & 93.9\% \\

 \bottomrule
\end{tabular}
 }
\vspace{-2mm}
\caption{\textit{ Sensitivity ($S_e$), Specificity ($S_p$), Accuracy ($Acc$) of the proposed method. $\dagger$: fine-tuned only on \app{}-data. *: fine-tuned on Keratron-data (stage-1) and 50\% of \app{}-data (stage-2). $\ddagger$: fine-tuned only on Keratron-data (stage-1).}}
\vspace*{-8mm}
\label{tab:result_comparison}
\end{table}
\smallskip\noindent\textbf{Results}: 
We evaluate the performance of our network on the task of keratoconus detection on heatmaps from the \app{}-data. 
In stage-2, the network is \textit{fine-tuned} on 50\% of \app{}-data and evaluated on the remaining 50\%; we perform this evaluation on 5 random splits, and report the average values over the 5 runs. 
We report sensitivity ($S_e = P_k/N_k$), specificity ($S_p = P_n/N_n$), and accuracy ($Acc = \frac{P_k+P_n}{N_k+N_n}$) values (Table~\ref{tab:result_comparison}), where $P_i, N_i$ are the number of correctly classified samples and total number of samples in class $i \in$ \{\textit{keratoconus}, \textit{non-keratoconus}\}, respectively. 
The final output of the dual-head CNN model is computed as the worst case possibility from the axial and tangential branch, \textit{i.e.}, if either of the branches predict the heatmap as keratoconus, we consider the final label as keratoconus.

Our network, comprising of dual-head CNN and 2-stage transfer learning, achieved an accuracy of 93.1\% ($S_e$=91.3\%, $S_p$=94.2\%) on the \app{}-data (Table~\ref{tab:result_comparison}). 
The 2-stage transfer learning showed significant improvement compared to traditional transfer learning of training only on the \app{}-data, which achieved an accuracy of 71.9\% ($S_e$=65.2\%, $S_p$=76.5\%).
Analogous to the \app{}-data, we also had access to the heatmaps generated by Keratron for the same set of 64 patients. Our network achieved an accuracy of 93.9\% ($S_e$=94.7\%, $S_p$=93.4\%) when evaluated on these Keratron heatmaps (without the stage-2 of fine-tuning).
In comparison, the Keratron-generated PPK-value based classification achieved an accuracy of 92.2\% ($S_e$=89.4\%, $S_p$=94.2\%). 
Additionally, we train an SVM on sim-K quantitative values, where the input vector comprises of \textit{[sim-k1, sim-k2, (sim-k1 -- sim-k2), (sim-k1 + sim-k2)/2]}. This classifier achieved a high specificity (100.0\%) but performed poorly in terms of sensitivity (80.4\%) (Table~\ref{tab:result_comparison}). This further shows the superiority of the proposed network.

\begin{figure}[t]
\begin{center}
    \centering
    \includegraphics[width=1\linewidth]{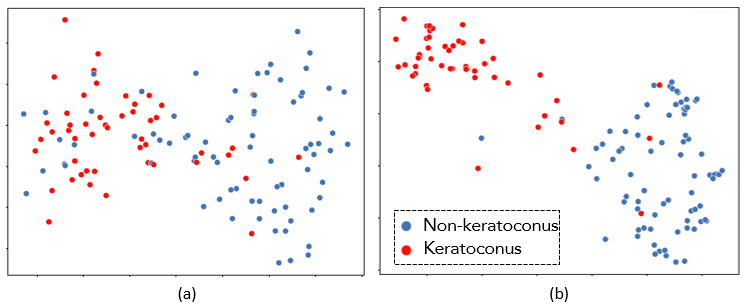}
\end{center}
\vspace{-5mm}
    \caption{\textit{t-sne plots (a) before training, and (b) after training, depicts a good separation of the learned features for keratoconus vs non-keratoconus test samples. (Note: This is shown for the fully-connected features for axial heatmaps; the tangential heatmap features show a similar separation.)}} 
    \label{fig:tsne}
\vspace{-4mm}
\end{figure}

Our proposed network is successful in learning discriminative features for the detection of keratoconus, as visible from the t-sne visualizations~\cite{tsne} (
Figure~\ref{fig:tsne}) of the learned representations from the second fully-connected layer (FC2 in Figure~\ref{fig:overview}) before and after training it for the axial heatmaps. (Note: the tangential heatmaps show a similar separation).

Furthermore, we analyze the effect of data augmentations and fine-tuning, by conducting an ablation analyses. Since the traditional transfer learning model performed poorly ($S_e$=65.2\%, $S_p$=76.5\%) due to the small size of \app{}-data, for the ablation experiments, we consider the model after stage-1 fine-tuning on Keratron-data as our base model.

\begin{table}[t]
\centering
 \resizebox{\linewidth}{!}{
\begin{tabular}{| c | c | c c c|}
\toprule
 \textbf{Fine-tuning} & \textbf{Augmentations} & \textbf{Se} & \textbf{Sp} & \textbf{Acc.}\\
\cmidrule(lr){1-1}
\cmidrule(lr){2-2}
\cmidrule(lr){3-5}

Stage-1 & None & 88.8\% & 10.5\% & 42.1\% \\
Stage-1 & Mixup & 84.7\% & 43.1\% & 59.9\% \\
Stage-1 & Dom. Spec. & 82.1\% & 39.9\% & 56.9\% \\
Stage-1 & Dom. Spec. + Mixup & 89.1\% & 65.0\% & 74.7\% \\
Stage-1,2 & None & 82.1\% & 85.6\% & 84.1\% \\
Stage-1,2 & Dom. Spec. + Mixup & 91.3\% & 94.2\% & 93.1\% \\
 \bottomrule
\end{tabular}
 }
\vspace{-2mm}
\caption{\textit{Ablations for fine-tuning, and domain-specific and mixup data augmentations. Evaluation is performed on the \app{} test split.}.}
\vspace*{-5mm}
\label{tab:ablations}
\end{table}

\smallskip\noindent\textbf{Data Augmentations Ablation}: 
Table~\ref{tab:ablations} demonstrates the utility of adding domain-specific data augmentations. Without adding any augmentations, the network performed poorly with an accuracy of 42.1\% ($S_e$=88.8\%, $S_p$=10.5\%). Adding \textit{mixup} or domain-specific (rotation, translation, scaling) augmentations increased the accuracy to 59.9\% and 56.9\%, respectively. On combining both types of augmentations, the accuracy further improves to 74.7\% ($S_e$=89.1\%, $S_p$=65.0\%).

\begin{figure}[t]
\begin{center}
    \centering
    \includegraphics[width=1\linewidth]{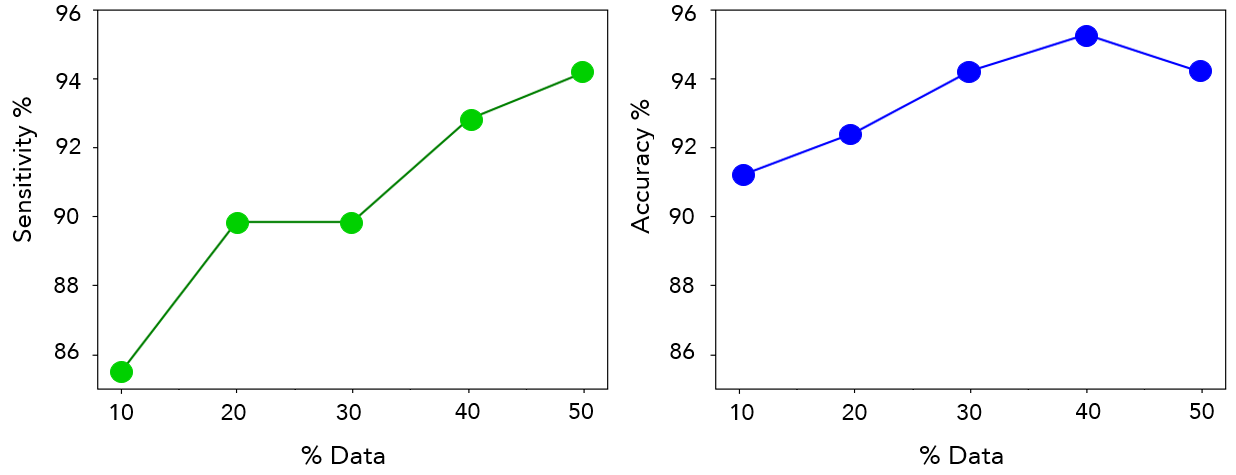}
\end{center}
\vspace{-5mm}
    \caption{\textit{Effect of \% data used for stage-2 fine-tuning on the \app{}-data.}} 
    \label{fig:fine_tune}
\vspace{-4mm}
\end{figure}

\smallskip\noindent\textbf{Fine-tuning Ablation}: Despite the improvements in performance after adding data augmentations, the performance of the network is not satisfactory.
Fine-tuning on a subset of \app{}-data (stage-2) drastically improved performance, with the network achieving an accuracy of 84.1\% ($S_e$=82.1\%, $S_p$=85.6\%) (Table~\ref{tab:ablations}) even without any data augmentations. When combined with data augmentations, the network accuracy improved to 93.1\% ($S_e$=91.3\%, $S_p$=94.2\%). The amount of \app{}-data used for fine-tuning impacts performance (Figure~\ref{fig:fine_tune}). Increasing the data for stage-2 fine-tuning from 10\% to 50\% improves accuracy by 3.1\%, thus hinting that more data can further improve the performance of our network. 
We found the sensitivity to increase uniformly with more data, however the accuracy decreased slightly on increasing stage-2 fine-tuning data from 40\% to 50\% (Figure~\ref{fig:fine_tune}). This can be attributed to our usage of a weighted loss function, which penalizes the network more for misclassifying keratoconus samples, and also owing to the fact that the number of non-keratoconus samples (68) is much larger than keratoconus samples (46).


\section{Conclusion}

In this work, we demonstrated the efficacy of a deep neural network to detect keratoconus using the curvature heatmaps generated by \app{}, a low-cost smartphone-based corneal topographer. We propose a dual-head CNN (one each for the axial heatmap and the tangential heatmap), and develop a 2-stage transfer learning strategy to train our network on a small-sized dataset (114 samples). 
We achieve results that are at par with doctors \cite{smartkc}, indicating that our proposed system can be used for diagnosis of keratoconus in practice. 
Although this work focuses on classifying corneal topography heatmaps, the proposed 2-stage transfer learning strategy can be applied generally to similar scarce data settings.
In future, to improve the performance of the network and increase robustness of our system, we plan to collect more data for training the network. Finally, we aim to deploy the \app{} system with the proposed novel CNN-based keratoconus classifier to aid in mass screening of keratoconus in schools, villages, and remote locations. This can help in early diagnosis of keratoconus and timely treatments, thus playing a crucial role in preventing needless blindness.




{
\small
\printbibliography
}
\end{document}